\documentclass[aps,prl,twocolumn]{revtex4-1}
\usepackage[utf8]{inputenc}
\usepackage{hyperref}
\usepackage{graphicx}  
\usepackage{dcolumn}   
\usepackage{bm}        
\usepackage{amssymb,amsmath}   
\usepackage{uri}

\usepackage[x11names, rgb, html]{xcolor}
\definecolor{sbase03}{HTML}{002B36}
\definecolor{sbase02}{HTML}{073642}
\definecolor{sbase01}{HTML}{586E75}
\definecolor{sbase00}{HTML}{657B83}
\definecolor{sbase0}{HTML}{839496}
\definecolor{sbase1}{HTML}{93A1A1}
\definecolor{sbase2}{HTML}{EEE8D5}
\definecolor{sbase3}{HTML}{FDF6E3}
\definecolor{syellow}{HTML}{B58900}
\definecolor{sorange}{HTML}{CB4B16}
\definecolor{sred}{HTML}{DC322F}
\definecolor{smagenta}{HTML}{D33682}
\definecolor{sviolet}{HTML}{6C71C4}
\definecolor{sblue}{HTML}{268BD2}
\definecolor{scyan}{HTML}{2AA198}
\definecolor{sgreen}{HTML}{859900}
\hypersetup{
  colorlinks = true,
  citecolor = {sred},
  urlcolor = {sblue}
}

\newcommand{\arcsinh}{\text{arcsinh}}

\begin{document}
\title{Dissipation bounds all steady-state current fluctuations}
\author{Todd R. Gingrich}
\email{toddging@mit.edu} 
\author{Jordan M.~Horowitz}
\author{Nikolay Perunov}
\author{Jeremy L. England}
\affiliation{Physics of Living Systems Group, Department of Physics, Massachusetts Institute of Technology, 400 Technology Square, Cambridge, MA 02139}

\begin{abstract}
Near equilibrium, small current fluctuations are described by a Gaussian with a linear-response variance regulated by the dissipation.
Here, we demonstrate that dissipation still plays a dominant role in structuring large fluctuations arbitrarily far from equilibrium.
In particular, we prove a linear-response-like bound on the large deviation function for currents in Markov jump processes.
We find that nonequilibrium current fluctuations are always more likely than what is expected from a linear-response analysis.
As a small-fluctuations corollary, we derive a recently-conjectured uncertainty bound on the variance of current fluctuations.
\end{abstract}

\maketitle

One of the most useful insights into thermodynamics has been that fluctuations near equilibrium are completely characterized by just one principle, the fluctuation-dissipation theorem~\cite{Kubo2012}.
Far from equilibrium, however, fluctuations exhibit less universal structure.
As such, characterizing the rich anatomy of nonequilibrium fluctuations has been handled on a case by case basis, with few universal nonequilibrium principles.
Notable exceptions are the fluctuation theorems~\cite{Kurchan1998, Crooks1999, Lebowitz1999, Andrieux2007, Chetrite2008, Jarzynski2011}, as well as fluctuation-dissipation theorems for nonequilibrium steady states~\cite{Harada2005, Speck2006, Baiesi2009, Prost2009, Seifert2010}.
Recently, Barato and Seifert have proposed a new kind of nonequilibrium principle, a thermodynamic uncertainty relation that expresses a trade-off between the variance of current fluctuations and the rate of entropy production~\cite{Barato2015}.
It reveals that away from equilibrium, dissipation continues to regulate small fluctuations.
While the thermodynamic uncertainty relation was not proven in general, analytical calculations and numerical evidence support its validity~\cite{Barato2015}.
Applications appear myriad, and already include insights into chemical kinetics as well as biochemical sensing~\cite{Barato2015Fano, Barato2015Dispersion}.

In this paper, we demonstrate that dissipation in fact constrains all current fluctuations.
In particular, we prove a pair of general thermodynamic inequalities for the large deviation function of the steady-state empirical currents in Markov jump processes.
Such processes model a variety of scenarios, including molecular motors~\cite{Altaner2015}, chemical reaction networks~\cite{Qian2005,Schmiedl2007}, and mesoscopic quantum devices~\cite{Esposito2009}.
Our analysis reveals that far from equilibrium, current fluctuations are always more probable than would be predicted by linear response~\cite{Maes2007,Maes2008On}.
Remarkably, our relationship bounds even rare fluctuations (large deviations), and by specializing to small deviations we obtain the thermodynamic uncertainty relation.

\begin{figure}
\includegraphics[width=0.47\textwidth]{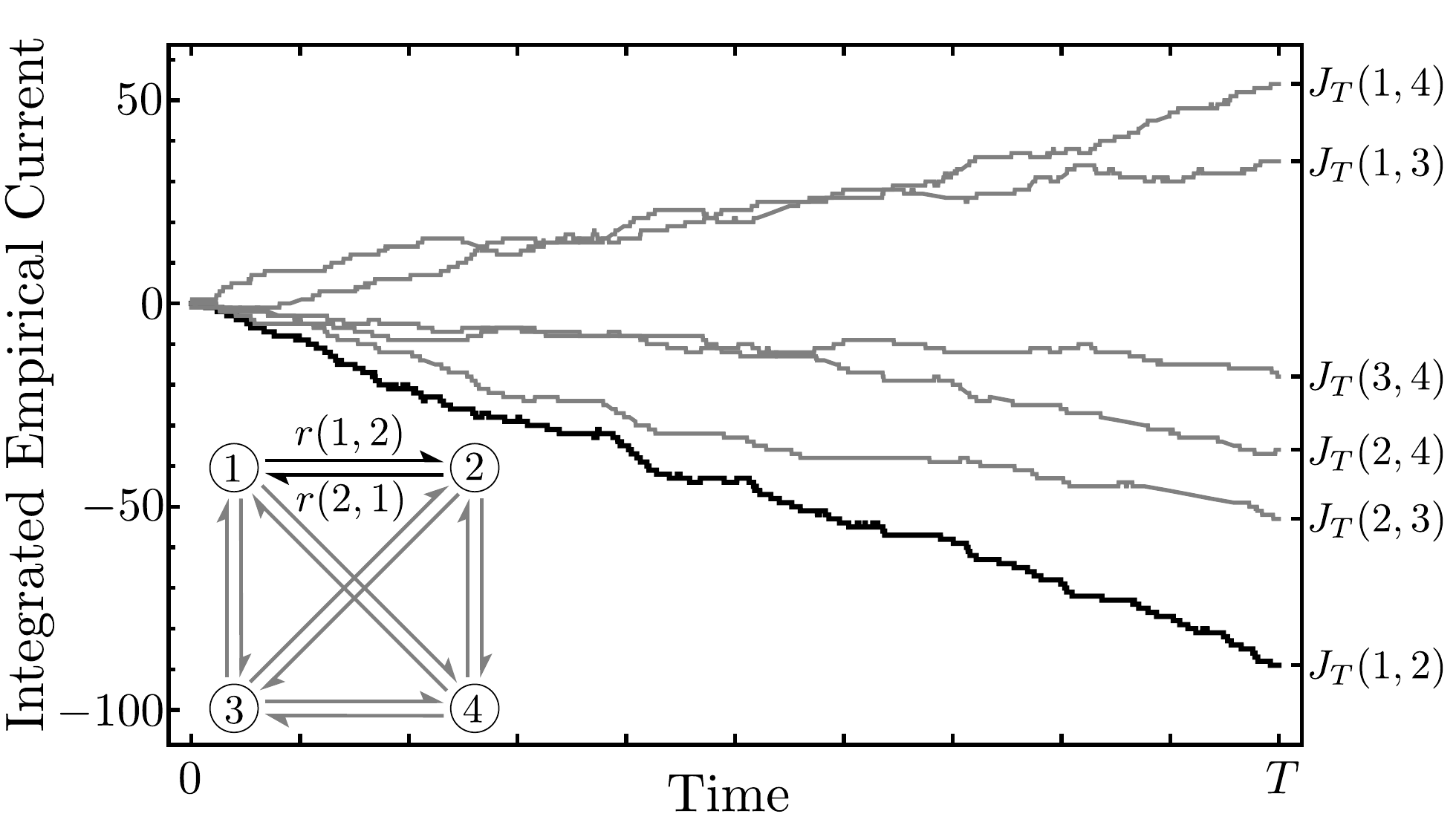}
\caption{Current fluctuations illustration: For a 4-state model (inset) in a nonequilibrium steady state, the integrated current $J_T$ -- net number of hops between pairs of states -- along each edge is plotted as a function of time.
Each integrated current displays an average rate perturbed by stochastic fluctuations.
}
\label{fig:4stateschematic}
\end{figure}

We have in mind a system with $N$ mesoscopic states (or configurations), $x=1,\dots, N$. 
Transitions between pairs of states, say from $y$ to $z$, are modeled as a continuous-time Markov jump process with rates $r(y,z)$~\cite{vankampen1992}.
It is convenient to picture these dynamics occurring on a graph (as in Fig.~\ref{fig:4stateschematic}), 
with vertices denoting states and edges (or links) symbolizing possible transitions.
We assume the dynamics to be ergodic and that $r(z,y) > 0$ whenever $r(y,z) > 0$, so the system's probability density relaxes to a unique steady state $\pi(x)$ in the long-time limit.
Thermodynamics enters by requiring the transitions to satisfy local detailed balance; the ratio of rates on each edge can then be identified with a generalized thermodynamic force
\begin{equation}
F(y,z) = \ln\left(\frac{\pi(y)r(y,z)}{\pi(z)r(z,y)}\right),
\end{equation}
which quantifies the dissipation in each transition~\cite{Esposito2010}.
For example, if a transition were mediated by a thermal reservoir at inverse temperature $\beta$, we have $F=\Delta s+\beta q$, where $\Delta s=-\ln[\pi(z)/\pi(y)]$ is the change in the system's stochastic entropy~\cite{Seifert2012} and $\beta q=\ln[r(y,z)/r(z,y)]$ is the heat dissipated into the reservoir.
Here and throughout, $k_{\rm B}=1$.

Now imagine watching the system evolve for a long time from time $t=0$ to $T$ as it jumps along a sequence of states $x(t)$, and we measure the \emph{integrated empirical current} through all the links by counting the net number of transitions along each edge,
\begin{equation}
J_T(y, z)\equiv \int_0^Tdt\, \left(\delta_{x(t^-), y} \delta_{x(t^+), z} - \delta_{x(t^-), z} \delta_{x(t^+), y}\right),
\label{eq:empiricalcurrent}
\end{equation}
where $x(t^\pm)$ denotes the state of the system just before and after a jump.
Its rate $j_T(y,z)=J_T(y,z)/T$, the empirical current, asymptotically converges in the infinite-time limit to the average steady-state value, $\lim_{T\to \infty} j_T(y, z) = j^\pi(y,z) \equiv \pi(y)r(y, z) - \pi(z)r(z, y)$.
For long finite times, fluctuations in the vector of empirical currents $j$ away from the typical value $j^\pi$ are possible, but exponentially rare, with a probability density that satisfies a large deviation principle, $P(J_T=Tj)\sim e^{-TI(j)}$~\cite{Touchette2009}.
The large deviation rate function $I(j)$ yields an extension of the central limit theorem, quantifying not just the Gaussian fluctuations about the typical value ($j^\pi$, which is the minimum of $I$), but also the relative likelihood of rare fluctuations.
In general, the determination of this large deviation function is challenging and analytical expressions are limited to particular models (e.g.~\cite{Harris2005, Derrida2007, Gorissen2012, Altaner2015}).

Our main result is a pair of general thermodynamic bounds on the rate function of empirical currents.
The first is a bound for the current fluctuations in terms of the rate of steady-state entropy production along each link $\sigma^\pi(y,z)=j^\pi(y,z)F(y,z)$:
\begin{equation}\label{eq:bound}
I(j)\le I_{\rm LR}(j)=\sum_{y<z}\frac{\left(j(y,z)-j^\pi(y,z)\right)^2}{4 j^\pi(y,z)^2}\sigma^\pi(y,z),
\end{equation}
where the sum extends over all edges.
As we will argue, such Gaussian fluctuations are what one would have expected from a linear-response analysis; as such, this bound is tightest within linear response.
The subscript ${\rm LR}$ reinforces this observation.
Further, the inequality is saturated not only at the minimum of $I$, $j= j^\pi$, but also at the symmetric point $j=-j^\pi$, as our bound also satisfies the fluctuation theorem for currents~\cite{Andrieux2007,Bertini2015Flows}. 

Our second inequality is a weakened form of  (\ref{eq:bound}) for any \emph{generalized current} expressed as a linear combination $j_{\rm d}=\sum_{y<z}d(y,z)j(y,z)$.
The key benefit of this weakened form is that now the current fluctuations are constrained by the average dissipation rate $\Sigma^\pi=\sum_{y<z}j^\pi(y,z)F(y,z)$, which is often easier to measure than the individual  entropy production rates $\sigma^\pi$: 
\begin{equation}\label{eq:dbound}
I(j_{\rm d})\le I_{\rm WLR}(j_{\rm d})=\frac{\left(j_{\rm d}-j^\pi_{\rm d}\right)^2}{4 (j^\pi_{\rm d})^2}\Sigma^\pi.
\end{equation}
The subscript WLR connotes a weakening of  (\ref{eq:bound}).
Our analysis reveals this bound is tightest, and indeed as strong as the linear-response bound, when $d=F$.
Under this condition, the generalized current is the dissipation rate, $\Sigma = \sum_{y<z} j(y,z)F(y,z)$:
\begin{equation}
I(\Sigma)\le I_{\rm WLR}(\Sigma) = \frac{\left(\Sigma-\Sigma^\pi\right)^2}{4 \Sigma^\pi}.
\label{eq:entropyproductionbound}
\end{equation}

Derivations of these inequalities appear at the end of this paper; here, we examine their meaning and explore their consequences.

Foremost, we stress that these bounds are not quadratic truncations of $I$ for small currents, but originate in a linear-response expansion about small force or entropy production. 
We can see this, roughly, by analyzing linear-response fluctuations in the entropy production rate $\sigma(y,z)=j(y,z)F(y,z)$.
Near equilibrium, the typical fluctuations are known to be Gaussian with a variance that is twice the mean $\sigma^\pi$~\cite{Speck2004, Maes2007},
\begin{align}
I_{\rm LR}(\sigma) &= \sum_{y<z}\frac{\left(\sigma(y,z)-\sigma^\pi(y,z)\right)^2}{4\sigma^\pi(y,z)}\\
 &= \sum_{y<z}\frac{\left(j(y,z)-j^\pi(y,z)\right)^2}{4j^\pi(y,z)}F(y,z).
\end{align}
This quadratic rate function indicates that in linear response each edge supports Gaussian current fluctuations with a variance regulated by the thermodynamic force.
Now imagine turning up the force, pushing the system away from linear response.
We would expect the typical currents to grow, due to both an increased bias and more activity in the number of jumps.
Remarkably, \eqref{eq:bound} implies that these effects are accompanied by an increase in the likelihood of rare current fluctuations, in excess of the linear-response prediction.

As a consequence of these bounds, we have a general proof of the conjectured thermodynamic uncertainty relation~\cite{Barato2015}.
Namely, the relative uncertainty in a generalized current $\epsilon^2_{\rm d}=\text{Var}(j_{\rm d})/(j^\pi_{\rm d})^2$, which is the variance normalized by the mean, verifies the inequality
\begin{equation}\label{eq:uncertainty}
\epsilon^2_{\rm d}\Sigma^\pi \ge 2.
\end{equation}
Thus, controlling current fluctuations by reducing their relative uncertainty $\epsilon^2_{\rm d}$ costs a minimal dissipation.
The inequality follows from  (\ref{eq:dbound}) by noting that the current's variance is obtained from the rate function as $I^{\prime\prime}(j_{\rm d}^\pi)=1/\text{Var}(j_{\rm d})$, and that the large-deviation inequality translates to the second derivative, since $I$ and $I_{\rm LR}$ have the same minimum.
A similar argument applied to  (\ref{eq:bound}) leads to an uncertainty relation for the current fluctuations along each transition, $\text{Var}(j(y,z))\ge 2j^\pi(y,z)^2/\sigma^\pi(y,z)$, which is tighter than the bound predicted using $d(u,v)=\delta_{(u,v),(y,z)}$ in~\eqref{eq:uncertainty}.
Furthermore, our derivation shows that these uncertainty relations are tightest in linear response and when $d\propto F$, as predicted by Barato and Seifert~\cite{Barato2015}. 

To illustrate the generalized-current bound  \eqref{eq:dbound} we numerically evaluate the rate function for two toy models: a 4-state model and the 1D asymmetric exclusion process (ASEP) with open boundary conditions.
\begin{figure}[t!]
\includegraphics[width=0.48\textwidth]{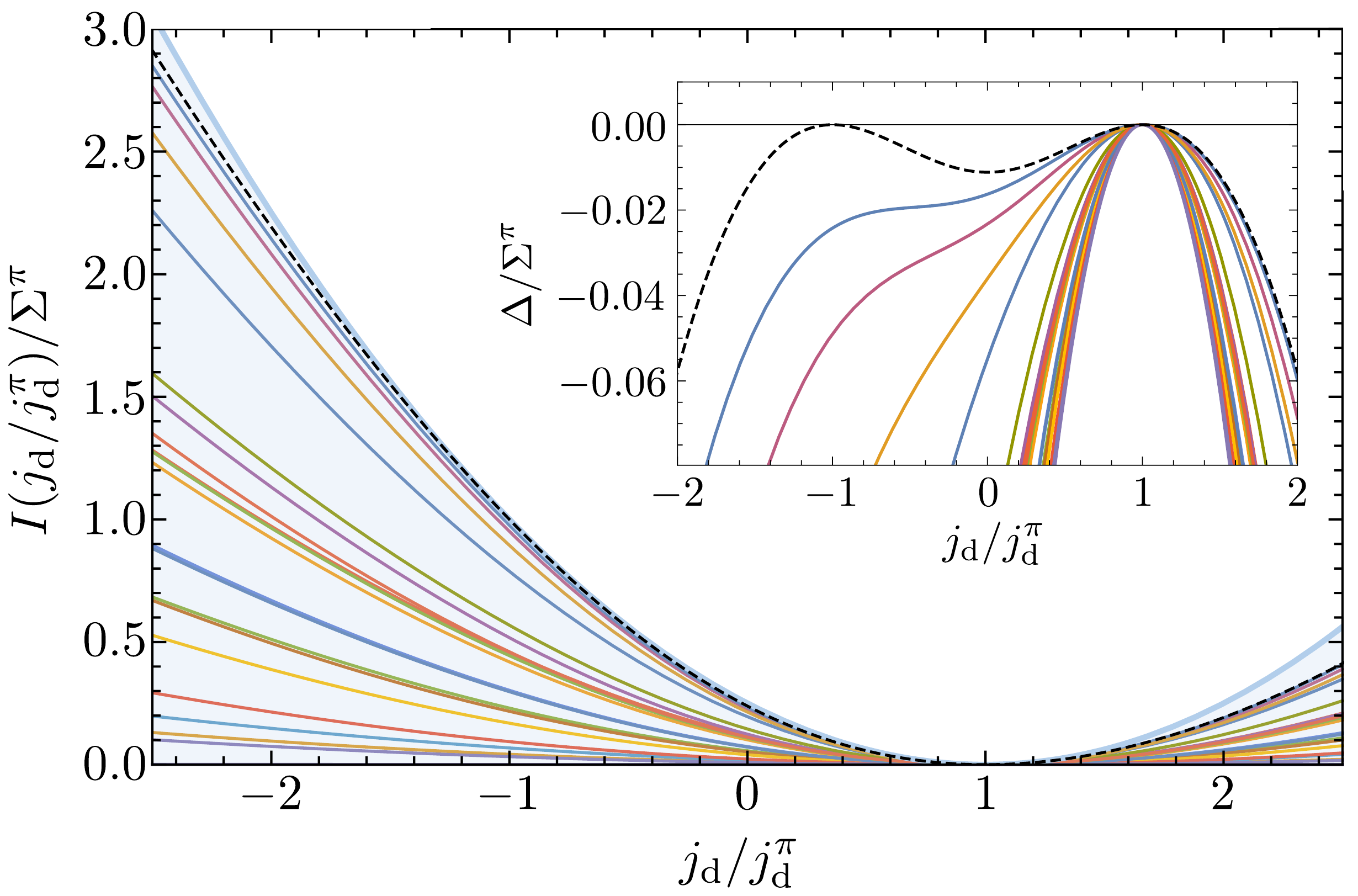}
\caption{Generalized-current fluctuations in the 4-state model:
Generalized currents $j_{\rm d}$ were constructed randomly by choosing $d(y,z)\in [-1, 1)$.
All the $I(j_{\rm d})$ (colored) and $I(\Sigma)$ (dashed black) fall in the blue-shaded region that satisfies the bound  (\ref{eq:dbound}), with the differences $\Delta \equiv I_{\rm WLR}(j_{\rm d}) - I(j_{\rm d})$ plotted in the inset.
Rates: $r(1,2)=3,\, r(1,3)=10,\, r(1,4)=9,\, r(2,1)=10,\, r(2,3)=1,\, r(2,4)=2,\, r(3,1)=6,\, r(3,2)=4,\, r(3,4)=1,\, r(4,1)=7,\, r(4,2)=9,\, r(4,3)=5$.
}
\label{fig:4state}
\end{figure}

Our first example is the multi-cyclic 4-state graph depicted in Fig.~\ref{fig:4stateschematic}.
The rate functions for the dissipation $\Sigma$ as well as for a collection of random generalized currents $j_{\rm d}$ were numerically computed using standard methods~\footnote{$I(j_{\rm d})$ is calculated as the Legendre transform of the scaled cumulant generating function $\psi(z) = \lim_{T\to\infty} \frac{1}{T} \ln \left<e^{z T j_d}\right>$, which was obtained numerically as the maximum (real) eigenvalue of a tilted rate matrix~\cite{Lebowitz1999, Lecomte2007, Touchette2009}} and plotted in Fig.~\ref{fig:4state}.
As required by our bound, all the  $I(j_{\rm d})$ fall below $I_{\rm WLR}(j_{\rm d})$ (within the blue-shaded region).
Interestingly, some generalized currents lie much closer to the bound than others.
In particular, the rate function for the dissipation $I(\Sigma)$ (black dashed) saturates the bound at $\Sigma =\pm \Sigma^\pi$.
Consequently, the  bound is significantly tighter for dissipation than for the other generalized currents, illustrating that the  tightness of the bound is quite sensitive to the choice of $d$.

\begin{figure}[t!]
\includegraphics[width=0.48\textwidth]{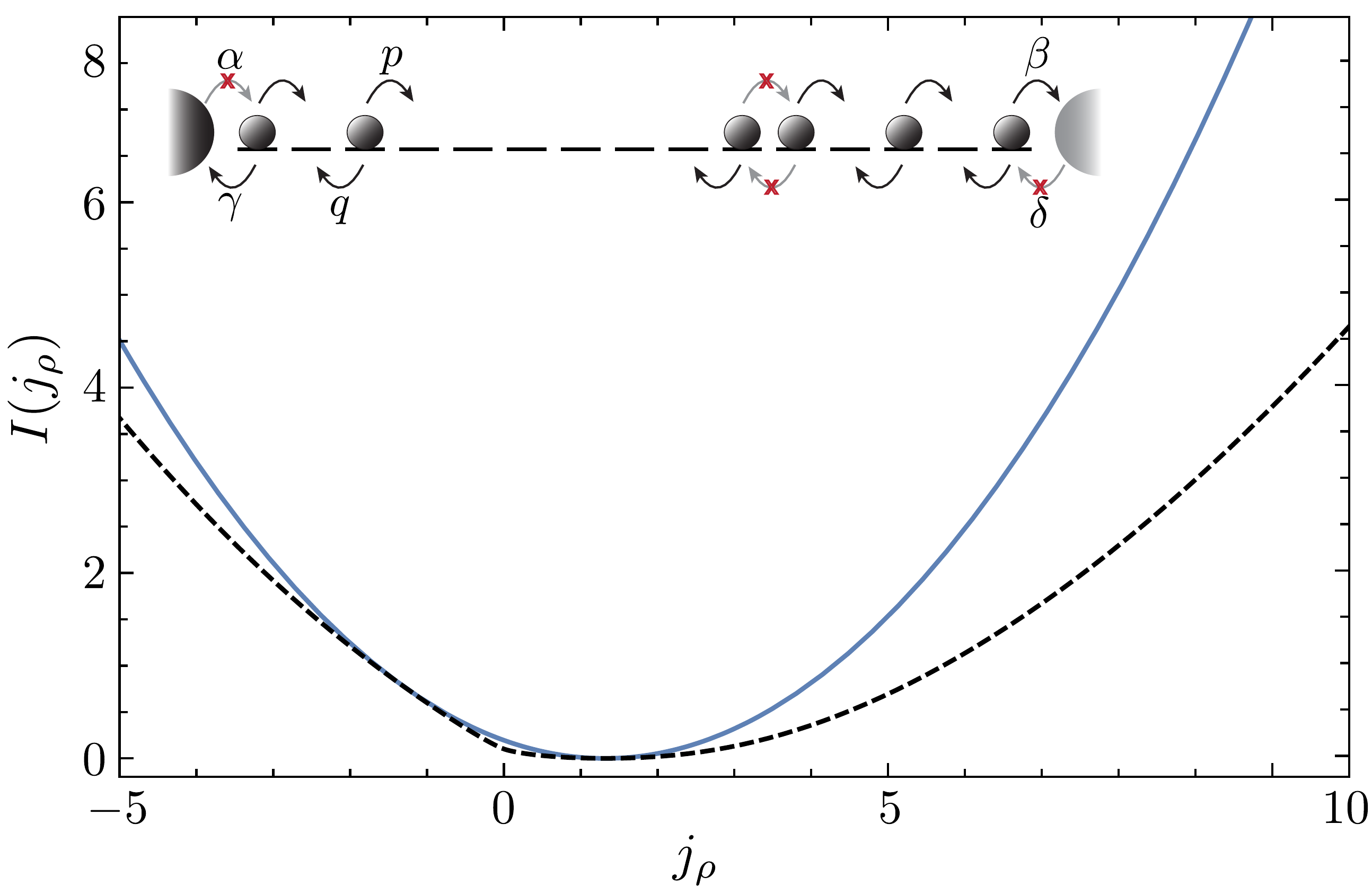}
\caption{ASEP Current fluctuations: Total current rate function $I(j_\rho)$ (black dashed) is bounded by $I_{\rm WLR}(j_\rho)$ (blue) for the $L=15$ ASEP.
Inset: Schematic of the $L=15$ ASEP with open boundaries.  Particles (shaded circles) jump to neighboring sites with rates $p,q$ in the bulk, and with rates given by Greek letters in/out of the boundary reservoirs (shaded semi-circles).
Blocked transitions are marked by a red ``x''.
Parameters, $\alpha = 1.25,\, \beta = 0.5,\, \gamma = 0.5,\, \delta = 1.5,\, p = 1,\, q = 0.5$, correspond to a high-density phase~\cite{Chou2011}.
}
\label{fig:asep}
\end{figure}

While the generalized-current bound (\ref{eq:dbound}) is not our strongest, we emphasize the important benefit that it avoids the computation of edgewise entropy production rates.
This advantage is especially profitable in many-particle dynamics.
To illustrate this point, we consider the current fluctuations in a canonical model of nonequilibrium particle transport, the 1D ASEP~\cite{Chou2011}.
The model consists of $L$ sites, occupied by at most one particle.
Particles hop into unoccupied neighboring sites with rates $p,q$, and enter or leave from two boundary particle reservoirs with rates $\alpha, \beta, \gamma, \delta$, as drawn in Fig.~\ref{fig:asep}.
The many-particle dynamics could be cast as a single-body dynamics on a graph, as in our first example, but the graph contains $2^L$ vertices and $(L+3)2^{L-2}$ edges.
For even moderately large $L$, it is impractical to record the average entropy production rates across all the edges, but we may more easily measure the mean dissipation $\Sigma^\pi$.
Indeed, ASEP only has one generalized force conjugate to the total particle current $j_\rho$ across the system: $\Sigma^\pi=j_\rho f_\rho$ with $f_\rho=\ln[\alpha\beta p^{L-1}/(\gamma\delta q^{L-1})]/(L+1)$.
This proportionality of $\Sigma^\pi$ and $j_\rho$ ensures that the WLR bound is equivalent to the stronger LR bound, explaining the tightness observed in Fig.~\ref{fig:asep}.

To summarize, dissipation constrains near-equilibrium current fluctuations, which in turn bound far-from-equilibrium fluctuations.
Thus, reducing current fluctuations carries a fundamental energetic cost.
This observation suggests a design principle: for fixed average dissipation and current, the fluctuations are most suppressed in a near-equilibrium process.
Such a principle may aid in engineering complex systems and understanding energy/accuracy trade-offs in biological physics~\cite{Lan2012}.
For instance, suppose we seek to construct a precise nonequilibrium process to reliably generate a current, e.g., a biochemical reaction network that produces a target molecule at a desired rate.
One can introduce energy-consuming metabolic cycles in an attempt to attenuate fluctuations.
However, with a fixed energy budget, it is impossible to surpass the linear-response bound, no matter how complex the design.

While finishing the present paper, we became aware of the preprint by Pietzonka, Barato, and Seifert~\cite{Pietzonka2015}, which conjectures~\eqref{eq:dbound} as a universal bound.  
Their work complements ours in that it provides additional analytical calculations for special cases and extensive numerical support.

{\it Derivation.---} To obtain (\ref{eq:bound}) and (\ref{eq:dbound}), we begin with the level 2.5 large deviations for continuous-time Markov processes~\cite{Barato2015Formal}.
This rate function describes the joint fluctuations for the empirical current $j_T$ with the empirical density
$p_T(y) \equiv \frac{1}{T}\int_0^T dt\, \delta_{x(t),y}$,
\begin{equation}
I(p, j) = \sum_{y<z} \Psi\left(j(y,z), j^p(y,z), a^p(y,z)\right),
\label{eq:level2.5}
\end{equation}
where 
\begin{equation}
\begin{split}
\Psi(j, \bar{j}, a) &= \left[j \left(\arcsinh\frac{j}{a} - \arcsinh\frac{\bar{j}}{a}\right) -\right. \\
&\left. \ \ \ \ \ \ \ \ \left(\sqrt{a^2 + j^2} - \sqrt{a^2 + \bar{j}^2}\right)\right],
\end{split}
\label{eq:level2.5psi}
\end{equation}
 $a^p(y,z) \equiv 2\sqrt{p(y)p(z)r(y,z)r(z,y)}$, and $j^p(y,z) \equiv p(y)r(y,z) - p(z)r(z,y)$ is the mean current associated with the empirical density~\cite{Maes2008, Bertini2015Large, Bertini2015Flows}.
This expression for $I(p,j)$ applies only for conservative currents, where $\sum_{z} j(y,z) = 0$ for all $y$; otherwise $I(p,j)$ is infinite.

We obtain (\ref{eq:bound}) in two steps: an application of the large-deviation contraction principle followed by a simple inequality.
First, we turn $I(p,j)$ into a rate function for just $j$ by using the contraction principle~\cite{Touchette2009}, which states that $I(j)=\inf_p I(p,j)$.
We can then bound the infimum by evaluating $I(p,j)$ at any normalized density $p(x)$.
The interesting choice is the steady state $\pi(x)$,
\begin{equation}\label{eq:pibound}
I(j)\le  \sum_{y<z} \Psi\left(j(y,z), j^\pi(y,z), a^\pi(y,z)\right).
\end{equation}
Next, we bound each $\Psi$ with a quadratic
\begin{equation}
\Psi(j,\bar{j},a) \le \frac{(j - \bar{j})^2}{4\bar{j}^2}(2{\bar j}\arcsinh\frac{\bar{j}}{a}),
\label{eq:quadraticboundedge}
\end{equation}
which can be verified by confirming that the difference between the right- and left-hand sides reaches its minimal value of zero at $j = \pm \bar{j}$.
We arrive at \eqref{eq:bound} by finally recognizing that $\sigma^\pi=2j^\pi\arcsinh(j^\pi/a^\pi)$.

Armed with this derivation, we can more clearly identify the physical interpretation of $I_\text{LR}$ as a linear-response bound.
We expand the contribution of each edge to~\eqref{eq:pibound} in terms of small force $F$ by utilizing the relationship $a^\pi = j^\pi/ \sinh(F/2)$,
\begin{equation}
\Psi(j, j^\pi, a^\pi) = \frac{(j - j^\pi)^2}{4 j^\pi} F - \frac{(j^2 - (j^\pi)^2)^2}{192(j^\pi)^3} F^2 + \mathcal{O}(F^3).
\label{eq:psiexpansion}
\end{equation}
The first order term describes exactly the predicted Gaussian linear response fluctuations.
All higher order corrections must have a negative sum.

The generalized current inequality in  (\ref{eq:dbound}) follows from  (\ref{eq:bound}) by contraction.
First, let us introduce a notation for the inner product on the vector space of currents, $\langle f, g\rangle=\sum_{y<z}f(y,z)g(y,z)$.
With this notation the generalized current is $j_{\rm d}=\langle d, j\rangle$, and the current conservation constraints can be  expressed by defining $h_k(a,b) = \delta_{ka} - \delta_{kb}$, so that $\langle h_k, j\rangle = 0$.
We thus have, by the contraction principle,
\begin{align}
I(j_{\rm d})&=\inf_j\{I(j)|\langle d, j\rangle=j_{\rm d}, \langle h_k, j\rangle = 0\ \forall k\} \\
& \leq \inf_j\{I_{\rm LR}(j)| \langle d,j \rangle=j_{\rm d}, \langle h_k, j\rangle = 0\ \forall k\},\label{eq:infJd}
\end{align}
since the infimum respects inequality.
As $I_{\rm LR}$ is a quadratic form with $N+1$ linear constraints, the minimization can be performed analytically; the solution is complicated and is presented below.
Inequality (\ref{eq:dbound}), however, follows readily once we observe that an upper bound on the infimum in \eqref{eq:infJd} can be obtained by evaluating $I_{\rm LR}$ at any $j=j^*$ that satisfies the constraints. 
We choose $j^* = (j_{\rm d}/j_{\rm d}^\pi) j^\pi$, which being proportional to the steady-state current must be conservative and trivially satisfies $\langle d, j^*\rangle = j_{\rm d}$.
Evaluating $I_{\rm LR}(j^*)$ gives the weaker quadratic bound,~\eqref{eq:dbound}.

 Finally, we demonstrate that (\ref{eq:dbound}) is indeed tightest when $d\propto F$ by minimizing \eqref{eq:infJd} directly with $N+1$ Lagrange multipliers to impose the constraints.
 The solution can be expressed compactly in terms of the pseudo-inverse of a symmetric square matrix with dimension $N+1$, $B_{kl} = \sum_{y<z} \frac{j^\pi(y,z)^2}{\sigma^\pi(y,z)} h_k(y,z) h_l(y,z)$ (where for notational convenience we have set $h_0 \equiv d$), as
\begin{equation}\label{eq:dboundPre}
I(j_{\rm d})\le I_{\rm LR}(j_{\rm d})= \frac{(j_{\rm d}-j_{\rm d}^\pi)^2}{4}B^{-1}_{00}.
\end{equation}
This inequality represents the LR bound contracted to the generalized scalar current $j_{\rm d}$.
$B$, which determines the values of the Lagrange multipliers associated to the constrained minimum, depends on the choice of generalized current.
Generally, $B^{-1}_{00}$ is onerous to compute, but it is simple in the special case when $d\propto F$.
Using the result, $B^{-1}_{00} = 1/\Sigma^\pi$, we find that the WLR and LR bounds for the entropy production fluctuations coincide, confirming the tightness of the WLR bound for entropy production.

This derivation sets the stage for exploring related aspects of nonequilibrium fluctuations, such as empirical density fluctuations~\cite{Donsker1975}, the impact of dynamical phase transitions~\cite{Garrahan2007}, and the role of the activity~\cite{Maes2008, Pietzonka2015}. 

\begin{acknowledgments}
We gratefully acknowledge Robert Marsland for helpful conversations.
We thank Raphael Chetrite and an anonymous referee for identifying an error in an earlier draft of this work.
This research is funded by the Gordon and Betty Moore Foundation to TRG as a Physics of Living Systems Fellow through Grant GBMF4513 as well as  JMH and JLE through Grant GBMF4343.
\end{acknowledgments}
\bibliography{refs}
\end{document}